\documentclass[aps,prl,preprint,superscriptaddress,nobibnotes,showkeys]{revtex4-2} 
\usepackage{graphicx,color}
\usepackage{amsmath}
\usepackage{bm}
\usepackage{dcolumn}
\usepackage{ifthen}
\usepackage{multirow}
\usepackage{threeparttable}
\usepackage{siunitx}
\usepackage{hyperref}
\usepackage{enumitem}
\usepackage{etoolbox}
\usepackage[utf8]{inputenc}
\usepackage[T1]{fontenc}
\hypersetup{pdfnewwindow=true, linkcolor=blue, colorlinks=true, anchorcolor=blue, citecolor=blue, filecolor=blue,  menucolor=blue, urlcolor=blue}
\usepackage{setspace}
\usepackage{newunicodechar}

\begin{document}
\title{MnBr$_2$ on the graphene on Ir(110) substrate: growth, structure, and super-moiré}

\author{Affan Safeer}
\email{safeer@ph2.uni-koeln.de}
\affiliation{II. Physikalisches Institut, Universit\"{a}t zu K\"{o}ln, Z\"{u}lpicher Stra\ss e 77, 50937 K\"{o}ln, Germany}
\author{Oktay G\"{u}lery\"{u}z} 
\affiliation{II. Physikalisches Institut, Universit\"{a}t zu K\"{o}ln, Z\"{u}lpicher Stra\ss e 77, 50937 K\"{o}ln, Germany}
\author{Nicolae Atodiresei} 
\affiliation{Peter Gr\"{u}nberg Institut (PG-1), Forschungszentrum J\"{u}lich, Wilhelm-Johnen-Stra\ss e, D-52428 J\"{u}lich, Germany}
\author{Wouter Jolie}
\affiliation{II. Physikalisches Institut, Universit\"{a}t zu K\"{o}ln, Z\"{u}lpicher Stra\ss e 77, 50937 K\"{o}ln, Germany}
\author{Thomas Michely} 
\affiliation{II. Physikalisches Institut, Universit\"{a}t zu K\"{o}ln, Z\"{u}lpicher Stra\ss e 77, 50937 K\"{o}ln, Germany}
\author{Jeison Fischer}
\affiliation{II. Physikalisches Institut, Universit\"{a}t zu K\"{o}ln, Z\"{u}lpicher Stra\ss e 77, 50937 K\"{o}ln, Germany}

\begin{abstract}

Single-layer MnBr$_2$ is grown on graphene (Gr) supported by Ir(110) and investigated using low-energy electron diffraction, scanning tunneling microscopy, and spectroscopy. The structure and epitaxial relationship with the substrate are systematically characterized. The growth morphology strongly depends on the growth temperature, evolving from fractal to dendritic and eventually to compact dendritic–skeletal islands, reflecting changes in the underlying surface diffusion processes. The pronounced variation in the apparent height with tunneling conditions for the magnetic insulator is explained based on the measured electronic density of states. MnBr$_2$ on Gr/Ir(110) constitutes a three-lattice system, giving rise to a super-moiré pattern -- a moiré of moirés. The super-moiré of MnBr$_2$/Gr/Ir(110) is unique, as it involves a virtual moiré of MnBr$_2$ with the Ir(110) surface lattice -- two lattices not in contact with each other. Using a careful Fourier analysis, the known properties of Gr/Ir(110), and the results of ab initio calculations, the origin of the virtual moiré is uncovered and related to the inhomogeneous binding of Gr to Ir(110). Comparative experiments with MnBr$_2$ on Gr/Ir(111) show similar growth and structure, but highlight the unique properties of the MnBr$_2$/Gr/Ir(110) super-moiré.  

\end{abstract}


\maketitle
\newpage

\section{Introduction}
Transition metal dihalides (TMDHs) are a class of materials that often exhibit magnetic ordering in their single-layer form owing to their unique electronic structures and strong spin-orbit coupling \cite{Botana2019, An2020, Bo2024, Kulish2017, McGuire2017, Basak2023}. Since the experimental discovery of intrinsic magnetic ordering in two-dimensional materials, single-layer TMDHs have attracted substantial research interest due to their potential as low-dimensional magnets and their applicability in spintronics and quantum technologies \cite{Huang2017, Elahi2022, Wang2022}. Beyond their intriguing magnetic properties, epitaxially grown single-layer TMDHs on graphene (Gr) undergo spontaneous self-trapping of charges, forming quasiparticles with extremely long lifetimes, identified as polarons \cite{cai2023, liu2023, Miao2025, Liang2025, Yao2024, Hao24, Song25}. Because these quasiparticles are observed in epitaxially grown single-layer TMDHs, understanding their exact nature requires a detailed analysis of the TMDH-Gr epitaxial relationship.

The investigation of single-layer TMDHs is often limited to first-principles calculations, with few experimental validations. An example lacking experimental realization is 2D manganese dibromide (MnBr$_2$). 
Bulk MnBr$_2$ crystallizes in a layered trigonal CdI$_2$-type structure (space group: $\rm{P\bar{3}m1}$), where the layers are separated by van der Waals (vdW) gaps. In each layer, Mn$^{2+}$ ions occupy an octahedral site coordinated by six Br$^-$ ions in an edge-sharing arrangement. At low temperatures, for bulk MnBr$_2$ a double-row striped antiferromagnetic ordering in the layers was measured \cite{Wollan58, farge1976, Iio1990, Sato1994}. 

Single-layer MnBr$_2$ has been investigated theoretically using density functional theory (DFT) \cite{Kulish17, Botana19, Luo20, Bo2024, Jesus25}. The different calculations agree that single layers are stable in the trigonal phase, similar to the structure of a single layer in bulk crystals. According to calculations, antiferromagnetic order is also present in the single-layer. The predicted band gap of the magnetic insulator ranges from 2.61\,eV \cite{Bo2024} to 4.21\,eV \cite{Luo20}. 

The preparation of single-layer MnBr$_2$ has not yet been realized. Preparation by exfoliation may be challenging because the strength of the in-plane bonds of TMDHs is weaker than that of materials such as Gr or transition metal dichalcogenides. Presumably for this reason, most experimental studies on single layers of TMDHs have been conducted using samples grown by molecular beam epitaxy (MBE) \cite{Zhou2020, Cai2020, Aguirre2024, Xiang2024, Hadjadj2023, liu2023, cai2023, Peng2020, Amini2024, wang2024, miao2023, Bikaljević2021, Terakawa2023, Cai2021}. Therefore, MBE of MnBr$_2$ is also the method of choice used in this study.  

In recent years, high-order moiré patterns that emerge from the interference of more than two lattices, so-called super-moiré or moiré of moiré patterns, have captured significant research interest within the condensed matter physics community because they provide an exceptional platform for realizing exotic quantum states \cite{Zhang2021, Ma2025, Kwan2025, Xie2025, Wang19, Andelkovic20, Hesp24, Lai2025, Craig24, Park25, Xia2025, Guerci2024}. For instance, super-moiré systems can host topological flat bands \cite{AlEzzi2024, Hao2024_flatband, Makov2024, Xia2025, Guerci2024} that enhance the electron-electron interactions, leading to the emergence of strongly correlated states, such as unconventional superconductors \cite{Balents2020}, fractional Chern insulators \cite{Liu2024}, quantum spin liquids \cite{Guangze2021}, and Wigner crystals \cite{Shayegan2022}. In $k$-space, a moiré results from an integer linear combination of vectors $\vec{k}[\mathrm{A}]$ and $\vec{k}[\mathrm{B}]$ belonging to two different lattices A and B, resulting in moiré vectors $\vec{k}[\mathrm{A/B}]$. Consequently, a super-moiré is considered an integer linear combination of two different sets of moiré vectors that originate from two different moirés \cite{Wang19, Andelkovic20}. This requires a minimal set of three lattices, A, B, and C, with sets $\vec{k}[\mathrm{A/B}]$ and $\vec{k}[\mathrm{B/C}]$. To date, super-moiré systems have been mostly limited to twisted triple-layer Gr \cite{Craig24, Park25, Xia2025, Xie2025} or heterostructures between Gr and hexagonal boron nitride \cite{Wang19, Andelkovic20, Hesp24, Lai2025, Ma2025}, where all the lattices have hexagonal symmetries. 

Here, we present a comprehensive characterization of single-layer MnBr$_2$, providing a foundation for future studies of its magnetic and polaronic properties. The investigation includes the determination of the structure and layer stacking, the analysis of growth or annealing temperature-dependent morphology linked to the underlying nucleation and diffusion processes, a systematic investigation of tunneling parameter-dependent layer height and its link to the electronic structure, and lastly, the determination of the super-moiré formed by the MnBr$_2$ layer with the substrate moiré. 

The case of MnBr$_2$/Gr/Ir(110) expands the concept and knowledge of super-moirés. Two of the three lattices are hexagonal (Gr and MnBr$_2$), but one is rectangular, namely the surface lattice of Ir(110).
The resulting super-moiré not only exhibits intriguing complexity, but is also formed differently than anticipated previously. Instead of resulting from the combination of moiré vectors $\vec{k}[\mathrm{A/B}]$ and $\vec{k}[\mathrm{B/C}]$, its dominating Fourier coefficient results from a vector $\vec{k}[\mathrm{A/C}]$, that is, in the present case $\vec{k}[\mathrm{MnBr_2/Ir(110)}]$. This is surprising since Ir(110) and the MnBr$_2$ layer are not in contact with each other. Hence, the effect is coined as virtual moiré. The presence of a virtual moiré in super-moiré formation is also unique, as highlighted by comparative experiments for MnBr$_2$ on Gr/Ir(111), which otherwise show a rather similar growth and structure of MnBr$_2$. 

\section{Results and discussion}
\subsection{Structure}
\begin{figure}[!ht]
\includegraphics[width=0.9 \textwidth]{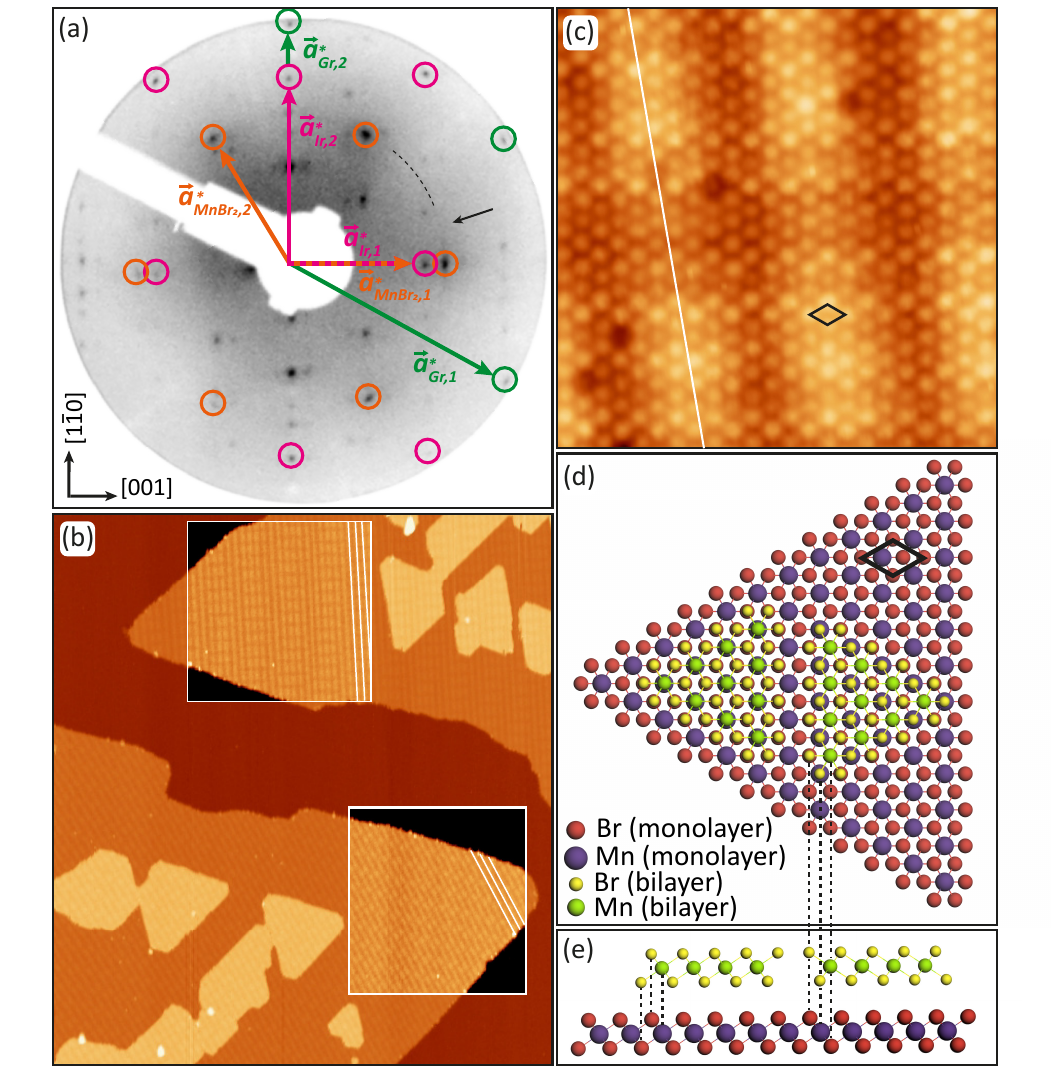}
\caption{Structural characterization of MnBr$_{2}$. (a) Contrast-inverted 90\,eV LEED pattern of MnBr$_{2}$ after the growth of 1.1\,ML at 400\,K on Gr/Ir(110). First-order Ir, Gr, and MnBr$_{2}$ reflections are encircled in magenta, green, and orange, respectively. Reciprocal Ir, Gr, and MnBr$_2$ primitive translations are indicated. (b) STM topography corresponding to the sample in (a). Contrast enhanced insets display striped moirés highlighted by white lines. (c) Atomically resolved STM image of MnBr$_{2}$. White line highlights moiré stripe, as in (b). (d) Top and (e) side view ball models of the structure and bilayer stackings of MnBr$_{2}$. All subfigures are oriented as indicated in (a), with the [001]-direction of Ir being horizontal. MnBr$_2$ unit cells are indicated as black rhombuses in (c) and (d). STM images are obtained at 300\,K with (b) $V_\mathrm{b}$ = -2\,V, 50\,pA and (c) $V_\mathrm{b}$ = -2\,V, 500\,pA. Image sizes: (b) 180\,nm $\times$ 200\,nm, and (c) 7.5\,nm $\times$ 7.5\,nm.}
\label{Fig_growth}
\end{figure}

The first insight into the MnBr$_{2}$ structure and epitaxial relationship with Gr/Ir(110) is provided through the low energy electron diffraction (LEED) pattern in Figure~\ref{Fig_growth}(a), taken after the growth of 1.1\,ML at 400\,K. The LEED pattern displays the first-order diffraction spots of the rectangular lattice of unreconstructed Ir(110) [encircled in magenta, with primitive reciprocal translations $\vec{a}^*_{Ir,1}$ and $\vec{a}^*_{Ir,2}$], the hexagonal Gr lattice [encircled in green, with primitive reciprocal translations $\vec{a}^*_{Gr,1}$ and $\vec{a}^*_{Gr,2}$], and the hexagonal MnBr$_{2}$ lattice [encircled in orange, primitive reciprocal translations $\vec{a}^*_{MnBr_2,1}$ and $\vec{a}^*_{MnBr_2,2}$]. Gr and Ir(110) are epitaxially aligned with a Gr zigzag direction parallel to $[001]_{\mathrm{Ir}}$ \cite{Kraus22}. MnBr$_2$ forms a hexagonal 2D lattice epitaxially aligned to Gr, with the two hexagonal lattices rotated by 30$^\circ$ with respect to each other. However, the epitaxy is not perfect, as manifested by the azimuthal smearing of the MnBr$_2$ diffraction spots and the presence of a faint, barely visible diffraction ring (highlighted by the segment of the dotted circle and the black arrow). The smearing of the diffraction spots becomes broader and the diffraction ring more intense when the growth temperature is lower. As typical in van der Waals epitaxy, the lower the growth temperature, the larger the orientation scatter. Using the Ir diffraction spots as a reference, analyzing multiple LEED patterns yields a MnBr$_{2}$ lattice parameter of 3.90\,Å $\pm$ 0.01\,Å. This value is in excellent agreement with both the experimental bulk MnBr$_{2}$ lattice constant of 3.873\,Å \cite{Ronda87} and the theoretically predicted values for single-layer MnBr$_{2}$ \cite{Bo2024,Jesus25,Luo20}. Other spots not mentioned and visible in the LEED pattern can be traced back to the moiré of Gr with Ir(110) \cite{Kraus22}.

Figure \ref{Fig_growth}(b) presents a scanning tunneling microscopy (STM) overview topograph of the same sample characterized by LEED, revealing two large first-layer MnBr$_{2}$ islands partially overlaid by smaller coalesced second-layer islands on top. The islands exhibit a triangular morphology with mostly straight edges, consistent with the C$_{3v}$ symmetry expected for a single layer of MnBr$_{2}$. At $V_\mathrm{b} = -2.0$\,V, the apparent height of the first-layer island is measured to be 4.15\,\AA~[see Figure S1 of the supporting information (SI)]. This value is lower than the expected geometric height of MnBr$_2$ on Gr, since the bulk interlayer spacing of MnBr$_2$ is 6.272\,\AA~\cite{Wollan58}. The underlying reason for this discrepancy is discussed in detail in the Section Apparent Height and Density of States. The contrast-enhanced insets of the first-layer islands in Figure \ref{Fig_growth}(b) also display different stripe patterns. The stripe patterns are the moirés of single-layer MnBr$_{2}$ with the underlying substrate, which are discussed in more detail in the Section Super-Moiré. The different orientations and appearances of the two stripe patterns result from the different orientations of the MnBr$_2$ islands consistent with the smeared LEED spots and the faint diffraction ring of MnBr$_2$. The islands are rotated by $\approx 6^\circ$ with respect to each other. 

Taking a closer look at Figure \ref{Fig_growth}(b), it is obvious that the first-layer MnBr$_{2}$ islands point in nearly opposite directions, where the triangular tip of the island in the upper part of the image points to the left and that in the lower part points to the right. The nearly opposite orientation of the large islands is consistent with the three-fold symmetric MnBr$_2$ growing on six-fold symmetric Gr. The fact that the orientations are not exactly opposite is consistent with the azimuthally smeared MnBr$_2$ reflections in the LEED pattern in Figure \ref{Fig_growth}(a).

Figure \ref{Fig_growth}(c) shows an atomically resolved STM image of single-layer MnBr$_{2}$, which confirms a hexagonal lattice that is consistent with the LEED results. In addition to the atomic-scale corrugation of $\approx$ 0.2\,\AA, the image exhibits nanoscale modulations. Most prominent are bright nearly vertical stripes, which are the same as those visible in the contrast-enhanced insets of Figure \ref{Fig_growth}(b). A small number of dark spots of atomic size are present in the atomically resolved image. Since their concentration does not show a dependence on growth or annealing temperature, they could result from a slight Br loss during the molecular sublimation in the Knudsen cell at 670\,K and be Br vacancies.  

The second-layer islands in Figure \ref{Fig_growth}(b) have two orientations with respect to the first-layer islands. Either their triangular envelope aligns with their base or opposes it. From the analysis of many STM topographs, it is found that aligned and opposing envelopes of second-layer islands are present in nearly equal proportions. As visualized in the ball model of Figure 1(d), the aligned envelopes correspond to bulk MnBr$_{2}$ which crystallizes in the CdI$_{2}$ structure \cite{Wollan58}, also denoted 1T-MnBr$_2$. In 1T-MnBr$_2$, layers composed of octahedrally coordinated Mn ion planes sandwiched between Br ion planes are stacked exactly on top of each other. The envelopes of opposite orientations indicate a stacking fault. Deviations from 1T stacking are well known for bulk MnBr$_2$: above 623\,K MnBr$_2$ in the CdCl$_2$ structure is the stable phase, that is, the layers are rhombohedrally stacked, giving rise to a tripled c-axis parameter \cite{Schneider1992}. This implies small free energy differences for variations in stacking, consistent with the formation of stacking faults under the non-equilibrium conditions of MBE growth. In agreement with our observations, DFT calculations in the BiDB database \cite{Pakdel24} found identical energies for the two experimental MnBr$_2$ stackings within the precision of DFT, that is, an energy difference of only 0.1\,meV\,/\AA$^2$.

\begin{figure}[!htb]
  \includegraphics[width=\textwidth]{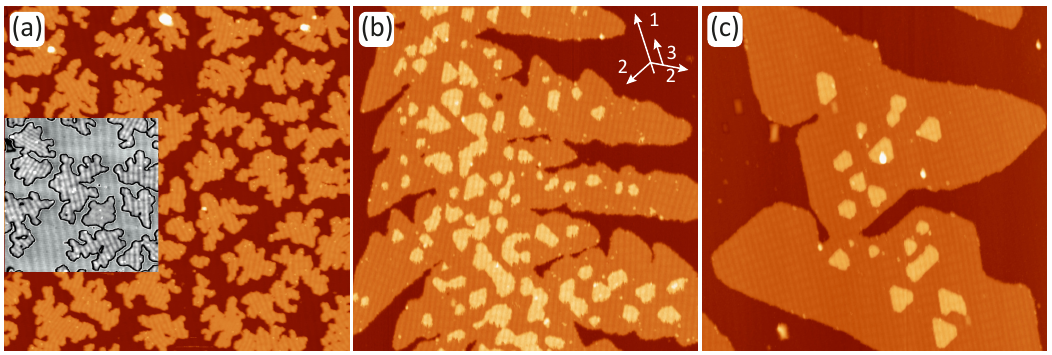}
  \caption{(a)-(c) STM topographs of MnBr$_{2}$ on Gr/Ir(110) after deposition of $\approx 0.6$\,ML at 80\,K, 300\,K, and 400\,K, respectively. The sample in (a) was briefly annealed to $\approx 150$\,K during transfer to the STM and then imaged at 77\,K. Inset in (a) uses double gray scale to make the moiré on Gr/Ir(110) and the moiré on the MnBr$_2$ islands simultaneously visible. Images in (b) and (c) were acquired at 300\,K. The white arrows in (b) indicate the dendritic growth directions of the primary, secondary, and ternary dendrite branches. STM imaging parameters: (a) $V_\mathrm{b} = -1$\,V and $I_\mathrm{t} = 50$\,pA, (b,c) $V_\mathrm{b} = -2$\,V and $I_\mathrm{t} = 50$\,pA  Image sizes: 160\,nm $\times$ 160\,nm.
  }
  \label{Fig_dep_temp}
\end{figure}
\subsection{Morphological Evolution with Temperature in Growth and Annealing}
Figure \ref{Fig_dep_temp} compares the morphology of MnBr$_2$ as a function of growth temperature. Figure \ref{Fig_dep_temp}(a), taken at 77\,K after growth at 80\,K with unavoidable annealing to $\approx 150$\,K in between due to sample transfer, reveals fractal island growth with smooth edges and no preferential orientation. The smooth edges suggest limited edge diffusion during the annealing period. In the absence of annealing at $\approx$ 150\,K, the islands would probably have retained an even more irregular fractal shape. As shown in the inset of Figure \ref{Fig_dep_temp}(a), most of the islands exhibit a single well-defined moiré orientation, except when grown together; an example of this is found in the upper left corner of the contrast-enhanced inset. One island in the inset does not display a visible moiré, for which the reason will be discussed in Section Super-moiré. The inset also shows the moiré of the Gr/Ir(110) substrate, which is always present, but much better visible for the specific bias used to take Figure~\ref{Fig_dep_temp}(a). The average moiré stripe periodicity is 3.3\,nm \cite{Kraus22}, although due to the magnification effect of the Gr/Ir(110) moiré for strain and defects, substantial variations occur around this average. In any case, even when the MnBr$_2$ moiré stripes are parallel to those of the substrate, their wavelength is substantially smaller. Second-layer islands are largely absent. The average distance between islands of about 25\,nm is much smaller than the typical separation between steps or other surface defects of about 150\,nm. This implies the dominance of homogeneous nucleation.  

Figure \ref{Fig_dep_temp}(b), acquired at 300\,K after growth at 300\,K, shows a segment of a large dendritic island. While the main visible branch of the dendrite grows from bottom to top [direction labeled 1 in the inserted sketch in Figure \ref{Fig_dep_temp}(b)], secondary branches grow away from it [directions labeled 2], maintaining angles of approximately 120$^\circ$. These angles correspond to the preferred growth directions, which are consistent with the three-fold symmetry of the crystal structure. The island edges are smooth on a much larger length scale than in Figure \ref{Fig_dep_temp}(a) and display undulations. Sharp grooves were observed between the secondary branches. Their origin can be traced back to the limited molecule supply from the diffusion field, which is preempted by the rapidly growing secondary branches and islands in the second layer. Some of the grooves are pinched, causing holes. The onset of tertiary branches is visible in Figure \ref{Fig_dep_temp}(b) [directions marked 3]. From the dendritic appearance, the rather smooth edges and the simultaneous existence of grooves, it must be concluded that step edge diffusion is effective on length scales of the order 5\,nm, but limited on large length scales of the order 50\,nm. Searching the sample by moving the STM scan frame shows dendritic growth all over, but the substrate steps and defects give rise to heterogeneous nucleation, making a quantitative assessment of the nucleation density impossible to achieve. Qualitatively, the strong reduction in the island number density with increasing growth temperature is consistent with classical nucleation theory \cite{Michely04}. A large number of second-layer islands nucleated on top of the dendritic island. This observation implies that the diffusion of MnBr$_2$ molecules on the first-layer MnBr$_2$ islands is hampered compared to diffusion on Gr/Ir(110), likely due to a larger activation barrier for molecule migration. Second-layer islands are seen to be reduced in number density next to the steps of the first-layer islands. This implies that for molecules diffusing on a first-layer island, the edge acts as a sink for molecules by incorporating them. Consequently, if there is any Ehrlich-Schwoebel barrier \cite{Ehrlich66,Schwoebel66,Michely04} for the descent of molecules from the first-layer island, it must be very small. 

Figure \ref{Fig_dep_temp}(c) taken at 300\,K and grown at 400\,K shows two first-layer islands grown together. The right island is a good example of dendritic-skeletal growth \cite{Michely04}. Its envelope is triangular, but its triangular sides are concave. The diffusion field has the strongest concentration gradient at the triangle tips; thus, the tips receive the largest supply of molecules. Although diffusion is efficient, transport along the edge is not fast enough to keep the edges straight, causing concave edges. Due to the higher growth temperature, the nucleation density of second-layer MnBr$_2$ islands is smaller compared to the 300\,K case. The zone denuded of second-layer islands adjacent to the first-layer island edges has grown in width, consistent with enhanced molecule diffusion at higher temperatures. The second-layer islands are compact and, if not coalesced, of hexagonal to triangular shape, since, due to their small size, diffusion is efficient in keeping the edges straight.

Taken together, the inferred picture of island nucleation and shape evolution is consistent with a strong increase with the growth temperature of (i) molecule diffusion on Gr/Ir(110) and (ii) step edge diffusion. The much higher diffusion coefficient of molecules on Gr/Ir(110) compared to that on the first-layer MnBr$_2$ islands implies that homogeneous nucleation has already ceased for the first-layer islands at 300\,K, while for the second-layer islands, it remains homogeneous throughout the investigated temperature range. 

Figure S2 of the SI for 400\,K growth of MnBr$_2$ on Gr/Ir(111) shows qualitatively the same behavior as for the 400\,K growth on Gr/Ir(110). Nucleation, second-layer nucleation, diffusion, and island shapes of MnBr$_2$ appear to be similar on both substrates. 

\begin{figure}[!ht]
  \includegraphics[width=0.5\textwidth]{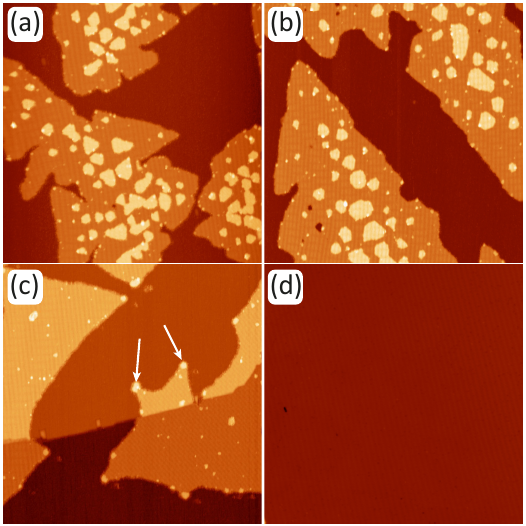}
  \caption{(a)-(d) The MnBr$_2$ sample of Figure 2(b) grown at 300\,K is annealed in isochronal steps of 120\,s successively to 370\,K, 470\.K, 570\,K, and 670\,K, respectively. White arrows in (c) highlight two pinning centers. Imaging is conducted 300\,K with $V_\mathrm{b}$ = -2\,V, $I_\mathrm{t} = 50$\,pA, Images sizes: 160\,nm $\times$ 160\,nm.}
  \label{Fig_anneal_temp}
\end{figure}

Figure \ref{Fig_anneal_temp} presents an isochronal annealing sequence of the sample shown in Figure \ref{Fig_dep_temp}(b) grown at 300\,K. The first two annealing steps to 370\,K [Figure \ref{Fig_anneal_temp}(a)] and 470\,K [Figure \ref{Fig_anneal_temp}(b)] do not cause significant morphological changes. The island edges tend to become smoother, and at 470\,K, the deep grooves of the dendrite branches disappear. Annealing to 570\,K as represented by Figure \ref{Fig_anneal_temp}(c) shows significant morphological changes. The second-layer islands disappeared. First-layer island steps show bright spots connected by concave-step segments. These observations indicate the onset of MnBr$_2$ molecule evaporation from the island edges. Due to their small radii of curvature, second-layer islands have the highest vapor pressure and disappear first. The bright spots at the first-layer island edges are pinning centers, presumably Mn clusters resulting from MnBr$_2$ decomposition. The evaporation of MnBr$_2$ molecules between the pinning centers causes these step segments to become concave, thereby lowering the vapor pressure at the edges. After annealing at 670\,K (Figure \ref{Fig_anneal_temp}(d)), no MnBr$_2$ remains on the surface, indicating that complete evaporation to the gas phase has occurred. This result is consistent with the Knudsen cell temperature of 670\,K used for the sublimation of MnBr$_2$.

\subsection{Apparent Height and Density of States}
\begin{figure}[!ht]
\includegraphics[width=0.5\textwidth]{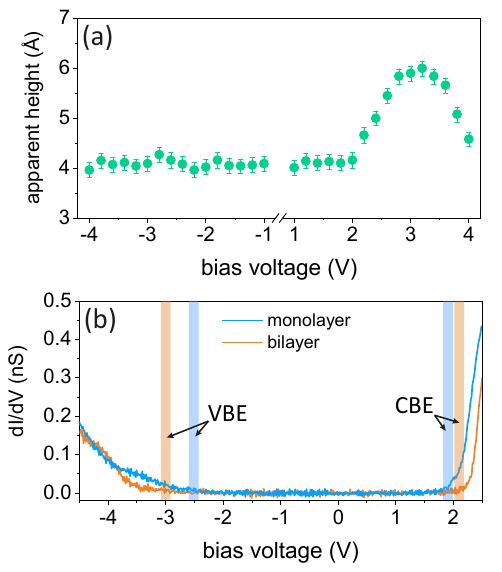}
\caption{(a) Apparent height of single-layer MnBr$_{2}$ with respect to Gr for $I_\mathrm{t} = 20$\,pA plotted as a function of $V_\mathrm{b}$. (b) Constant-height $\mathrm{d}I/\mathrm{d}V$ spectra of single-layer and bilayer MnBr$_{2}$. The spectra are obtained with $V_{st}$ = 2.5 V, $I_{st}$ = 100\,pA, $f_{mod}$ = 667\,Hz, and $V_{mod}$ = 20\,mV.}
\label{fig_STS and height}
\end{figure}
The apparent STM height of single-layer MnBr$_2$ on Gr/Ir(110) is plotted in Figure~\ref{fig_STS and height}(a) as a function of the bias voltage $V_\mathrm{b}$. Over a large range from -4\,V to 2\,V it is rather insensitive to $V_\mathrm{b}$ with an almost constant value of $\sim$4\,\AA. Only above $V_\mathrm{b} = 2\,V$ it increases, reaches at $V_\mathrm{b}$ = 3.2\,V a maximum of 5.95 $\pm$ 0.05\,\AA\ and then decreases again. The maximal measured height at $V_\mathrm{b}$ = 3.2\,V almost reaches the experimental c-axis lattice parameter of bulk MnBr$_2$, which is 6.272\,\AA~\cite{Wollan58}. We note that the precise apparent heights measured depend on $I_\mathrm{t}$ and the tip state, but its $V_\mathrm{b}$-dependence, as represented in Figure~\ref{fig_STS and height}(a) is generic.  

Also for second-layer islands the apparent heights are in the range of 4\,\AA~- 6\,\AA, dependent on $V_\mathrm{b}$, $I_\mathrm{t}$, and the tip state. For an example height profile of a second-layer island, see Figure~S1 of the SI.
The situation is similar to that of CoCl$_2$, where different apparent STM heights ranging from 4.8\,\AA~to 8.0\,\AA~and  strong bias dependencies were reported \cite{liu2023,cai2023,Hao24}. Atomic force microscopy provides a height of 6.3\,\AA~for CoCl$_2$, reasonably close to the expected bulk value of 5.9\,\AA~\cite{Kerschbaumer25}. As discussed in the following, electronic effects are responsible for the variation in the apparent height measured by STM.
\\
Figure~\ref{fig_STS and height}(b) displays a differential conductance ($\mathrm{d}I/\mathrm{d}V$) spectrum acquired from a defect-free region more than 10\,nm from the edges of single-layer MnBr$_{2}$ (blue curve). The differential conductance is, in the first approximation, proportional to the local density of states of the sample. 
The valence band edge (VBE) is located at $V_\mathrm{b} = (-2.5 \pm 0.1)$\,V and the conduction band edge (CBE) at $V_\mathrm{b} = 1.9 \pm 0.1$\,V.
The resulting total gap of $(4.4 \pm 0.2)$\,eV makes MnBr$_2$ an insulator. 
\\
Screening through the Gr substrate may be expected to have renormalized our measured band gap to values smaller than of  freestanding MnBr$_2$, as was observed for instance for MoSe$_2$ \cite{Ugeda14} or MoS$_2$ \cite{vanEfferen22}. Consistent with this speculation, we observe a larger band gap for bilayer MnBr$_2$, for which screening through the Gr substrate is reduced. As shown in Figure~\ref{fig_STS and height}(b), for bilayer MnBr$_2$ (red data) the VBE is located at $V_\mathrm{b} = (-3.0 \pm 0.1)$\,V and the CBE at $V_\mathrm{b} = (2.1 \pm 0.1)$\,V. The total resulting gap of $(5.1 \pm 0.2)$\,eV is significantly larger than the band gap of the single-layer. A similar trend has been reported for CoCl$_{2}$ grown on graphite (see the Supporting Information of ref.~\citenum{liu2023}), where the bilayer displays a larger band gap than the single-layer. For freestanding single-layer MnBr$_2$ previous DFT calculations found a band gap of 4.21\,eV \cite{Luo20} or 3.95\,eV \cite{Jesus25} and an even smaller band gap for MnBr$_2$ bulk. These results are consistent with MnBr$_2$ being an insulator, but seem to underestimate the band gap.

Evidently, the strong variation in apparent height, as shown in Figure~\ref{fig_STS and height}(a), is related to the electronic structure of MnBr$_2$. In the range of the band gap between $V_\mathrm{b} = (-2.5 \pm 0.1)$\,V and $V_\mathrm{b} = (1.9 \pm 0.1)$\,V tunneling is between the tip and the moiré substrate Gr/Ir(110). The tunnel barrier is composed of the vacuum barrier between the tip and MnBr$_2$ and the barrier within MnBr$_2$ between its two surfaces facing vacuum and Gr/Ir(110). The barrier within MnBr$_2$ is defined by its CBE. Since the CBE is lower in energy than the vacuum level, the presence of MnBr$_2$ enhances the tunneling probability and thus gives rise to a non-zero height of MnBr$_2$ in the constant current mode. 

When $V_\mathrm{b}$ decreases below $-2.5$\,V, a new tunneling channel opens: electrons can then tunnel also out of the MnBr$_2$ valence band. However, it is far below the Fermi level of the sample. Compared to electrons close to the sample Fermi level, their effective tunneling probability is lower by far, because of a much larger effective barrier height. Thus, the effect of tunneling out of the deep valence band states $-2.5$\,eV below the Fermi level is neglectable compared to tunneling out of Gr/Ir(110) states near the Fermi level. Therefore, no significant change in the apparent barrier height is observed. 

When $V_\mathrm{b}$ increases above $ 1.9$\,V, tunneling into the MnBr$_2$ conduction band sets in, thus giving rise to a new tunneling channel with a low effective barrier height. To maintain the same tunneling current under constant current tunneling conditions, this causes the tip to withdraw and gives rise to an increased apparent height. The decrease in apparent height beyond $V_\mathrm{b} = 3.2$\,V indicates that above 3.2\,eV the density of states in MnBr$_2$ decreases again, as also evident in the DFT calculations of ref.~\citenum{Jesus25}. 

\subsection{Super-moiré}

\begin{figure}[!ht]
\includegraphics[width=\textwidth]{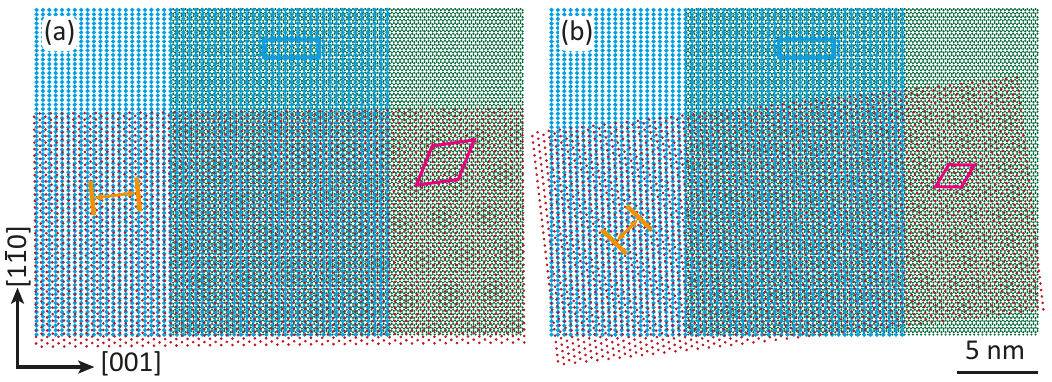}
\caption{Ball model of moiré formation through overlay of the Ir(110) surface lattice (starting left), the Gr lattice (starting right), and one Br sublattice of MnBr$_2$ (starting at the bottom). The MnBr$_2$ lattices are rotated counterclockwise in (a) by $0.5^\circ$ and in (b) by $6.5^\circ$ with respect to the Ir$[1\bar{1}0]$ direction. The moiré unit cells of Ir(110) with Gr and Gr with MnBr$_2$ are indicated by blue rectangles and pink rhombuses, respectively. The wavelength of the moiré resulting from the interference between Ir(110) and MnBr$_2$ is indicated by orange double arrows.}
\label{fig_Moiré_construction}
\end{figure}

The first insight into the super-moiré in the MnBr$_2$/Gr/Ir(110) system is obtained by overlaying ball-model representations of the three constituent lattices, as shown in Figure~\ref{fig_Moiré_construction}. The Ir(110) surface lattice (left) overlaps in the central area of both subfigures with the Gr honeycomb lattice (right). On these lattices, one Br sublattice (or equivalently the Mn sublattice) is overlaid (red balls, lattice shifted down) at two different angles. The angles of $0.5^\circ$ in Figure~\ref{fig_Moiré_construction}(a) and $6.5^\circ$ in Figure~\ref{fig_Moiré_construction}(b) for the dense-packed rows of the MnBr$_2$ lattice with respect to Ir$[1\bar{1}0]$ were selected to reproduce qualitatively the moiré stripes of the two islands in Figure \ref{Fig_growth}(b). On the left of each subfigure, a clean stripe moiré formed by Ir(110) with MnBr$_2$ is observed, in the upper middle the rectangular Gr/Ir(110) moiré (less pronounced), and on the right, a hexagonal moiré of Gr with MnBr$_2$ evolves, as indicated by an orange double arrow, a blue rectangle, and a pink rhombus, respectively. In the lower middle, where the three lattices overlap, the MnBr$_2$/Ir(110) moiré stripes survive, but become modulated due to the additional Gr lattice.

In fact, these stripes are the dominant features in wavelength and orientation in the moirés highlighted in Figure~\ref{Fig_growth}(b). When the angle between the dense-packed rows of the MnBr$_2$ lattice with respect to Ir$[1\bar{1}0]$ increases beyond $6.5^\circ$, the spacing of the stripes becomes smaller and reaches a minimum of 0.6\,nm at $30^\circ$. Due to the low growth temperature, the islands in Figure~\ref{Fig_dep_temp}(a) display an almost random orientation and consequently a large scatter in the orientation and width of the moiré stripes. The absence of a visible stripe pattern in one of the fractal islands in the inset of Figure~\ref{Fig_dep_temp}(a) is a consequence of a large angle near $30^\circ$, where the stripe pattern is not more resolved due to its small wavelength and the limited resolution on the scale of the topograph.

Although in the construction presented here, optically realized, it remains somewhat mysterious how a moiré of two lattices not in contact with each other, namely MnBr$_2$ and Ir(110), can form the dominant STM corrugation feature in the MnBr$_2$ islands.  

\begin{figure}[!ht]
\includegraphics[width=\textwidth]{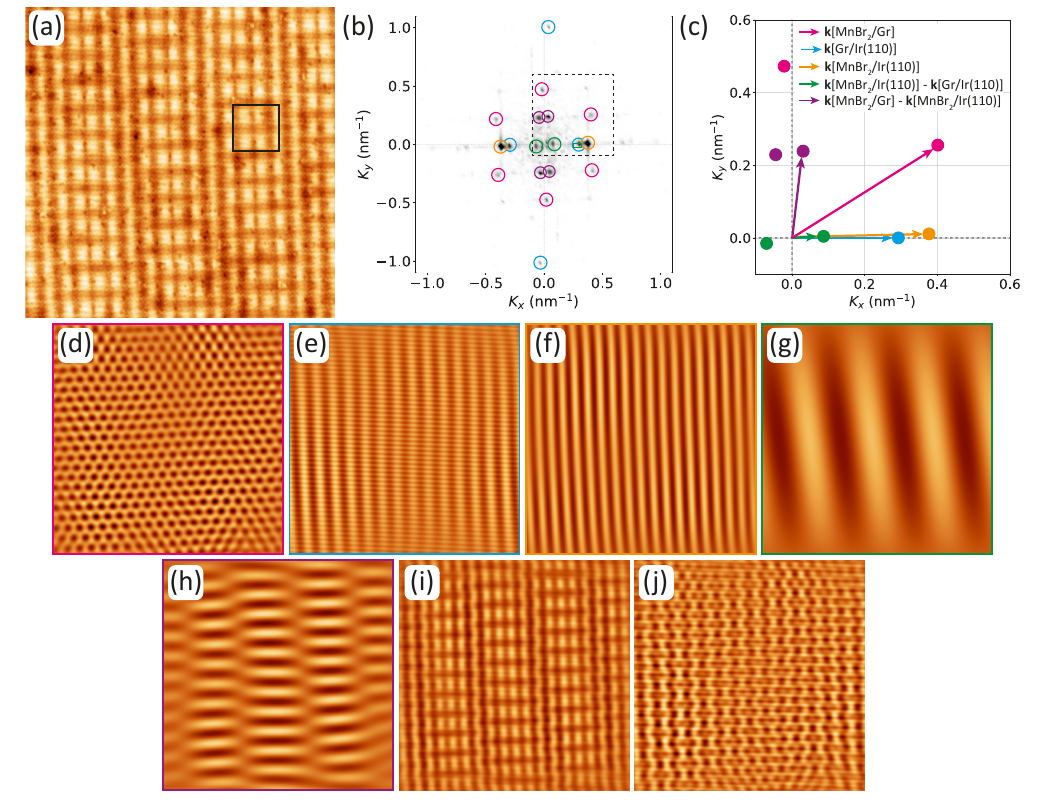}
\caption{Fast Fourier transformation (FFT) analysis of the MnBr$_2$/Gr/Ir(110) super-moiré. (a) High resolution STM topography of MnBr$_2$. Black box indicates size of atomic resolution topograph in Figure~\ref{Fig_growth}(b). (b) Inner part (small $\vec{k}$) of the FFT of the STM image shown in (a) with color-coded circles surrounding the spots. (c) Enlarged schematics of dashed box in (b) using same color code. The relevant $\vec{k}$ vectors are indicated. (d)-(h) are inverse FFTs using the spots in (b) encircled (d) pink, (e) blue, (f) orange, (g) green, and (h) purple. (i) is the inverse FFT using all spots encircled in (b), while (j) is the inverse FFT resulting from the spots encircled pink and blue. STM imaging parameters in (a) are $V_\mathrm{b} = -1$\,V and $I_\mathrm{t} = 50$\,pA. Image size is 50\,nm $\times$ 50\,nm.}
\label{fig_FFT_moiré}
\end{figure}

 The high resolution topography shown in Figure~\ref{fig_FFT_moiré}(a) exhibits additional complex moiré features beyond what can be visualized in a ball model. With an angle of about $0.5^\circ$ between the dense-packed MnBr$_2$ rows and the Ir$[1\bar{1}0]$ direction it corresponds to the situation visualized by the ball model in Figure~\ref{fig_Moiré_construction}(a) and highlighted in the upper inset of Figure \ref{Fig_growth}(b). To obtain a deeper understanding and to identify the Fourier components of the super-moiré, we performed a fast Fourier transform (FFT) analysis of Figure~\ref{fig_FFT_moiré}(a). 
The central part of the FFT limited to small $\vec{k}$ or large wavelengths is represented in Figure~\ref{fig_FFT_moiré}(b), where the relevant peaks are encircled in color. An enlarged and schematic version of the dashed box in Figure~\ref{fig_FFT_moiré}(b) is shown in Figure~\ref{fig_FFT_moiré}(c) using the same color code. The inverse FFT analysis of sets of equally colored peaks provides insight into the main components of the super-moiré. 

{When selecting only the FFT peaks encoded in pink forming a hexagon in Figure~\ref{fig_FFT_moiré}(b) with wave vectors like $\vec{k}\mathrm{[MnBr_2/Gr]}$, the inverse FFT results in Figure~\ref{fig_FFT_moiré}(d). The image represents the hexagonal moiré between  MnBr$_2$ and Gr already visible in the ball model of Figure~\ref{fig_Moiré_construction}(a) where a unit cell of the moiré is indicated by a pink rhombus. It has a lattice parameter of 2.5\,nm.}

When selecting only the four FFT peaks encoded in blue in Figure~\ref{fig_FFT_moiré}(b) with wave vectors like $\vec{k}\mathrm{[Gr/Ir(110)]}$, the resulting inverse FFT is shown in Figure~\ref{fig_FFT_moiré}(e). The image represents the rectangular moiré between Ir(110) and Gr already visible in the ball model of Figure~\ref{fig_Moiré_construction}(a) where a unit cell of the moiré is indicated by a blue rectangle. The orthogonal periodicities of the Gr/Ir(110) moiré are 1.0\,nm and 3.3\,nm \cite{Kraus22}. This moiré is also present between the MnBr$_2$ islands in the contrast-enhanced inset of Figure \ref{Fig_dep_temp}(a), with the 3.3\,nm well visible. The assignment is also supported by comparing the FFTs of plain Gr/Ir(110) and MnBr$_2$/Gr/Ir(110) as done in Figure S3 of the SI. Figure~\ref{fig_FFT_moiré}(e) also closely matches the topography of Gr/Ir(110) as shown in Figure S3(a). 

A traditional super-moiré would result from differences or sums of the corresponding moiré wave vectors, i.e., from combinations like
$\vec{k}\mathrm{[MnBr_2/Gr]} - \vec{k}\mathrm{[Gr/Ir(110)]}$. However, such Fourier components do not exist in the FFT. Instead, other
components are present.

When selecting the two most intense FFT peaks encoded in orange with wave vectors $\pm \vec{k}\mathrm{[MnBr_2/Ir(110)]}$, the resulting inverse FFT is shown in Figure~\ref{fig_FFT_moiré}(f). The image represents the interference between the MnBr$_2$ and Ir(110) lattices. The pattern has a row spacing of 2.7\,nm -- distinctly smaller than the 3.3\,nm periodicity of Gr/Ir(110) -- and gives rise to a pronounced line pattern in the MnBr$_2$/Gr/Ir(110) super-moiré. This line pattern is also clearly visible in Figure~\ref{fig_Moiré_construction}(a) on the left, where the Ir(110) surface rows overlap with the MnBr$_2$ lattice. Its periodicity is indicated in Figure~\ref{fig_Moiré_construction}(a) by an orange double arrow. As mentioned above, this virtual moiré of two lattices not in contact with each other dominates the STM corrugation in the MnBr$_2$ islands and gives rise to the stripes visible still in larger-scale topographs of MnBr$_2$ islands, that is, to the stripes on MnBr$_2$ islands highlighted in the insets of Figure \ref{Fig_growth}(b) and Figure~\ref{Fig_dep_temp}(a). 

When selecting only the FFT peaks encoded in green, the resulting inverse FFT is shown in Figure~\ref{fig_FFT_moiré}(g). These spots are due to the difference in the wave vector of the orange and blue spots, that is, from combinations like
$\vec{k}\mathrm{[MnBr_2/Ir(110)]} - \vec{k}\mathrm{[Gr/Ir(110)]}$ in Figure~\ref{fig_FFT_moiré}(b) and (c). The superposition of the two corresponding stripe patterns results in a beating pattern with a period of approximately 13.5\,nm. This modulation is visible in the STM topograph of Figure~\ref{fig_FFT_moiré}(a). 

Finally, when selecting only the FFT peaks encoded in purple, the resulting inverse FFT is shown in Figure~\ref{fig_FFT_moiré}(h). These spots are due to the difference in the wave vectors of the pink and orange spots, that is, from combinations like $\vec{k}[\mathrm{MnBr_2/Gr}] - \vec{k}[\mathrm{MnBr_2/Ir(110}]$. 

When selecting all FFT peaks encircled in Figure~\ref{fig_FFT_moiré}(b), the resulting FFT is shown in  Figure~\ref{fig_FFT_moiré}(i). It reproduces all relevant features of the complex super-moiré, as obvious from a comparison with the STM topograph of Figure~\ref{fig_FFT_moiré}(a). In order to reproduce the super-moiré, the MnBr$_2$/Ir(110) moiré is indispensable. The related wave vectors, like $\vec{k}[\mathrm{MnBr_2/Ir(110}]$, are contained in the orange, green, and purple spots. Selecting only the spots related to the MnBr$_2$/Gr and Gr/Ir(110) moirés (spots encircled pink and blue) does not provide an adequate description of the super-moiré as visible from Figure~\ref{fig_FFT_moiré}(j).  

\begin{figure}[!ht]
\includegraphics[width=0.5\textwidth]{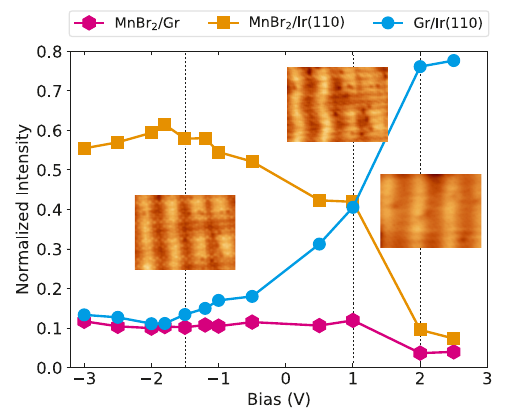}
\caption{Bias voltage dependence of the six pink,  four blue, and two orange Fourier peaks of Figure~\ref{fig_FFT_moiré}(b) as color-coded in Figures~\ref{fig_FFT_moiré}(b) and (c). These peaks are linked to wave vectors like $\vec{k}\mathrm{[MnBr_2/Gr]}$, $\vec{k}\mathrm{[MnBr_2/Gr]}$, and $\vec{k}\mathrm{[MnBr_2/Ir(110)]}$, respectively. The intensity of each component is normalized with the sum of all FFT peaks encircled in Figure~\ref{fig_FFT_moiré}(b). STM topographs in insets are 11\,nm $\times$ 8\,nm and taken at $V_\mathrm{b} = -1.5$\,V, $+1$\,V, and $+2$\,V as indicated by arrows and with $I_\mathrm{t} = 50$\,pA.}
\label{fig_FFT_bias}
\end{figure}

Additional insight into the origin of the super-moiré of MnBr$_2$ on Gr/Ir(110) is obtained from the analysis of the bias dependence of its Fourier components. Figure~\ref{fig_FFT_bias} plots the relative FFT intensities of the pink, orange and blue peaks of Figure~\ref{fig_FFT_moiré}(b) linked to the MnBr$_2$/Gr,  MnBr$_2$/Ir(110) and Gr/Ir(110) moirés, respectively. The pink and orange components are stable for $V_\mathrm{b} \leq +1$\,V with the orange component by far most intense. At  $V_\mathrm{b} = +2$\,V the components drop close to zero. The blue component is for negative   $V_\mathrm{b}$ on a low and quite stable level, increases at positive  $V_\mathrm{b}$ and jumps to new plateau for  $V_\mathrm{b} \geq +2$\,V, where it is the only component with significant intensity. The increase of the wavelength from 2.7\,nm (related to $\vec{k}\mathrm{[MnBr_2/Ir(110)]}$) to 3.3\,nm (related to $\vec{k}\mathrm{[Gr/Ir(110)]}$) with increasing bias voltage is visualized by the STM insets in Figure~\ref{fig_FFT_bias}.

The interpretation of these observations is straightforward. At positive bias with of $V_\mathrm{b} \geq +2$\,V tunneling is into the conduction band of MnBr$_2$. Thereby, the MnBr$_2$/Gr interface as well as the virtual MnBr$_2$/Ir(110) interface are invisible to the tunneling current. The experimental corrugation is due to the height variation of the Gr/Ir(110) substrate, which is imposed on the soft MnBr$_2$ layer. The geometrically imprinted Gr/Ir(110) moiré leaves the related blue Fourier component as dominant, while all other components vanish.

For $V_\mathrm{b} \leq 1.9$\,V tunneling is in the MnBr$_2$ band gap, that is, tunneling is \textit{through} MnBr$_2$ directly to the MnBr$_2$/Gr interface. This explains why the interface-related components are much larger than when tunneling into the conduction band. The tunneling current in this bias range is dominated by states around the Fermi level. Specifically at all negative bias voltages electrons at the Fermi level have the lowest tunneling barrier and thus dominate the tunneling current. This explains the relative constancy of the Fourier components in this bias range. As noted above when discussing the apparent height of single-layer MnBr$_2$, even when at $V_\mathrm{b} \leq -2.5$\,V tunneling out of the valence band becomes possible, this tunneling channel is insignificant due its large tunneling barrier.

\begin{figure}[hbt!]
\includegraphics[width=\textwidth]{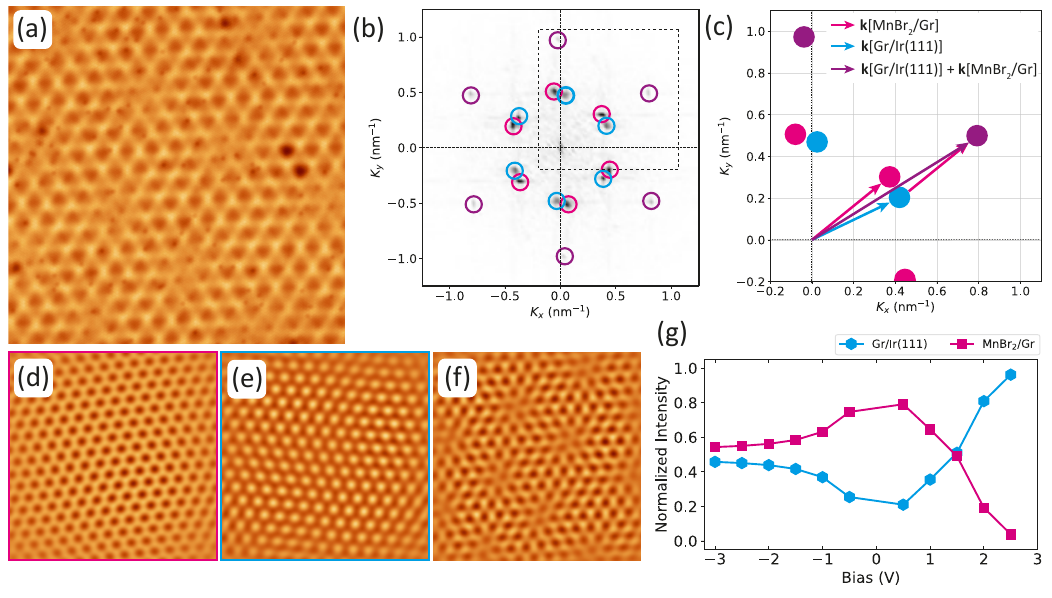}
\caption{FFT analysis of the MnBr$_2$/Gr/Ir(111) super-moiré. (a) Moiré resolution STM topography of MnBr$_2$. (b) Inner part (small $\vec{k}$) of the FFT of the STM image shown in (a) with color-coded circles surrounding the spots. (c) Schematic enlarged section of dashed box in (b) using same color code. The relevant $\vec{k}$ vectors are indicated. (d) and (e) are inverse FFTs using the spots in (b) encircled (d) pink and (e) blue, respectively. (f) is the inverse FFT using all spots encircled in (b) pink and blue. (g) Bias voltage dependence of the six pink and six blue Fourier peaks of (b). These peaks are linked to wave vectors like $\vec{k}\mathrm{[MnBr_2/Gr]}$ and $\vec{k}\mathrm{[Gr/Ir(111)]}$, respectively. The intensity of each component is normalized with the sum of all FFT peaks encircled (b). STM imaging parameters in (a) are   $V_\mathrm{b}$ = -2\,V, 100\,pA.  Image size is  35\,nm $\times$ 35\,nm.
} 
\label{fig_111} 
\end{figure}

The uniqueness of the MnBr$_2$/Gr/Ir(110) super-moiré is illuminated by comparing it to another three-lattice system forming a super-moiré. Figure~\ref{fig_111} investigates the case of MnBr$_2$/Gr/Ir(111), of which Figure~\ref{fig_111}(a) shows an STM topograph taken at the same $V_\mathrm{b}$ as Figure~\ref{fig_FFT_moiré}(a). It is dominated by the moiré of MnBr$_2$ with Gr [pink encircled components in Figure~\ref{fig_111}(b) with inverse FFT in Figure~\ref{fig_111}(d)]. The hexagonal Gr/Ir(111) moiré \cite{Busse11} with its 2.53 nm periodicity and 0.4\,\AA~corrugation is also present in the FFT, giving rise to the blue encircled components in Figure~\ref{fig_111}(b) with inverse FFT in Figure~\ref{fig_111}(e). The two moirés are slightly tilted with respect to each other. The super-moiré results from their interference. When the inverse FFT of the spots corresponding to the two moirés (encircled blue and pink), a hexagonal super-moiré with a periodicity of $\approx$13.2\,nm becomes obvious as is in the original STM topograph of Figure~\ref{fig_111}(e). As there are only few super-moiré unit cells present in the original STM topograph, the difference wave vectors $\vec{k}\mathrm{[MnBr_2/Gr]} - \vec{k}\mathrm{[Gr/Ir(111)]}$ are presumably too low in intensity to be visible in the FFT. From the analysis of Figure~\ref{fig_111} we conclude: (i) The FFT of the super-moiré shown in Figure~\ref{fig_111}(b) displays no virtual MnBr$_2$/Ir(111) component. (ii) The super-moiré is established by creating an inverse FFT only from the spots related to the moirés of MnBr$_2$/Gr, as shown in Figure~\ref{fig_111}(f). This is the expected way of super-moiré formation, but unlike the case of MnBr$_2$/Gr/Ir(110), where the corresponding inverse FFT in Figure~\ref{fig_FFT_moiré}(j) does not reproduce the super-moiré. (iii) The bias dependence of the interface Fourier component MnBr$_2$/Gr and of the geometric Fourier component Gr/Ir(111) shows a very similar behavior as the corresponding components of MnBr$_2$/Gr/Ir(110): below the CBE the interface component dominates the Fourier intensity, while in the conduction band only the geometric component is relevant [compare Figure~\ref{fig_111}(g)].   

The unique feature of the MnBr$_2$/Gr/Ir(110) is thus the existence and relevance of the virtual MnBr$_2$/Ir(110) moiré, that is, of the MnBr$_2$ and Ir(110) lattices that are not in contact with each other. Insight into the origin of this unique feature is provided by comparing Gr/Ir(110) with Gr/Ir(111).

Gr is physisorbed to Ir(111) with only a slight chemical modulation \cite{Busse11}. The C-Ir distance is for all C atoms larger than 3.2\,\AA, thus of a typical van der Waals distance \cite{Busse11, Hamalainen13}. There is no charge redistribution in or transfer to/from Gr that has the periodicity of the underlying Ir(111) surface lattice (see Figure 2 in ref.~\citenum{Busse11}). The Gr Dirac cone remains largely intact \cite{Pletikosic09}. The Ir(111) lattice cannot be detected by STM through the Gr cover.

\begin{figure}[!ht]
\includegraphics[width=\textwidth]{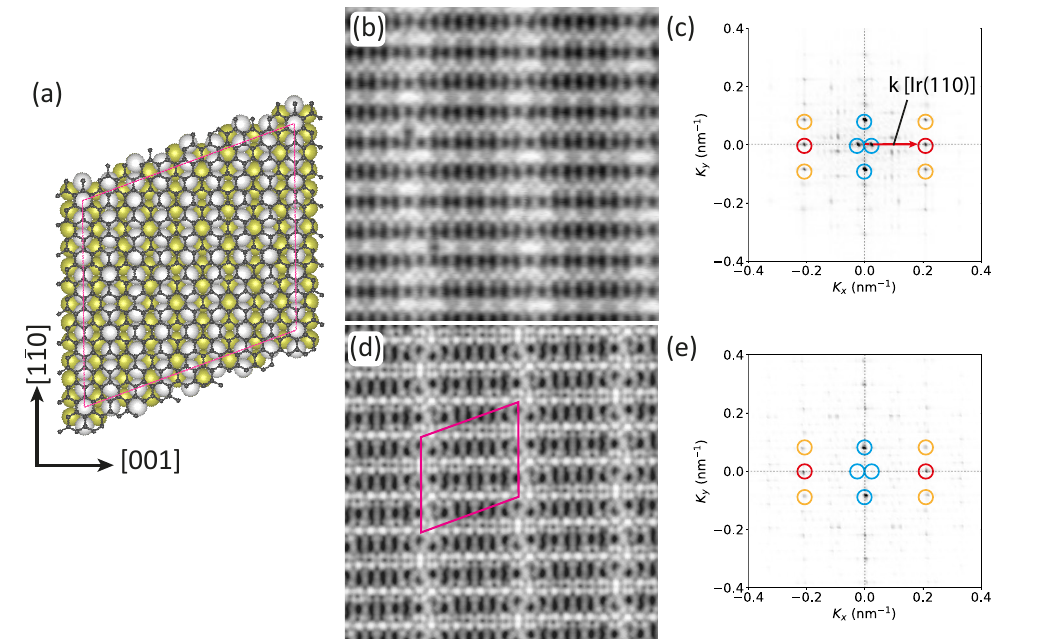}
\caption{Ir(110) rows visible through Gr on Ir(110). (a)  Ball model of Gr on Ir(110). Green gold balls: bottom-layer Ir atoms; white balls: top-layer Ir atoms; dark gray balls: C atoms. A pink rhombus indicates the supercell size in the DFT calculations. (b) STM topograph of Gr/Ir(110) moiré in high resolution. Image size 10\,nm $\times$ 10\,nm. $V_\mathrm{b} = -0.1$\,V and $I_\mathrm{t} = 1$\,nA. (c) FFT of (b). Moiré spots are encircled blue, spots due to Ir(110) dense-packed rows are encircled red, spots that are $\vec{k}$-vector sum of blue and red spots are encircled orange. (d) Simulated STM image for $V_\mathrm{b} = -0.1$\,V plotting the partial charge density in the energy interval $e V_\mathrm{b} = [-0.1\mathrm{eV}; 0]$ above the surface for an isosurface value of $4.2 \times 10^{-11}$\,e/\AA$^3$. 
(e) FFT of (d).}
\label{fig_charge}
\end{figure}

Gr is chemisorbed to Ir(110) with a pronounced modulation between strong and weak chemisorption \cite{Kraus22}. The C atoms above the dense-packed rows of the Ir(110) top layer [light beige in Figure~\ref{fig_charge}(a)] are strongly chemisorbed (distances of C-Ir down to 2.1\,\AA), while C atoms sitting in between the densely packed rows [above the furrows, yellow in Figure~\ref{fig_charge}(a)] are weakly chemisorbed (distances C-Ir up to 3.1\,\AA). There is strong charge redistribution and transfer of charge to Gr near the Fermi level with a periodicity of the underlying Ir(110) lattice (see Figures 3 and 4 of ref.~\citenum{Kraus22}). The Dirac cone is destroyed for Gr on Ir(110) with a substantial density of states, where it might be expected.

Figure~\ref{fig_charge}(b) is a high resolution STM topograph of the Gr/Ir(110) moiré with $V_\mathrm{b}= -0.1$\,V. In addition to the moiré itself, the periodicity of the dense-packed top layer rows of Ir(110) with their spacing of 3.84\,\AA~are well visible in the topograph. In the corresponding FFT of Figure~\ref{fig_charge}(c) the spots related to the Ir rows encircled in red are the most intense spots except of those of the moiré itself. Note also that linear combinations of the moiré spots $\vec{k}[\mathrm{Gr/Ir(110}]$ and the row spots with $\vec{k}[\mathrm{Ir(110)}]$ are intense (encircled orange). This implies that at the interface the bromine lattice of MnBr$_2$ with vectors $\vec{k}[\mathrm{MnBr_2}]$  is directly superimposed with the Ir(110) row spacing to form the virtual MnBr$_2$/Ir(110) moiré with wave vectors $\vec{k}[\mathrm{MnBr_2/Ir(110}]$. The presence of spots linked to the Ir(110) rows with $\vec{k}[\mathrm{Ir(110)}]$ and linear combinations of this $\vec{k}$-vector with those of the moiré are generic and observed for a large range of tunneling parameters. 

As example, for a tunneling resistance smaller by a factor of 20 compared to Figure~\ref{fig_charge}(b) also the Gr lattice is resolved (see Figure 1(c) of ref.~\citenum{Kraus22}). Nevertheless, the corresponding FFT shown in Figure S2 in the Supplemental Material of ref.~\citenum{Kraus22} still displays high intensity for the Ir(110) surface lattice. 

Our inferences are supported by DFT calculations. Figure~\ref{fig_charge}(d) shows the corresponding simulated STM topograph of Gr/Ir(110) which displays a pattern strikingly similar to the experimental STM topograph. In addition to the moiré pattern, the periodicity of the Ir(110) rows dominates the simulated STM topograph. This is underlined by the high intensity of the component encircled red in the corresponding FFT of Figure~\ref{fig_charge}(e). Also in a contour plot of the total charge density the Ir(110) row periodicity is clearly present as shown Figure~S4 of the SI. In conclusion the probability density of the wave functions of Gr/Ir(110) extending into the vacuum, where MnBr$_2$ is to be placed, carries a strong Fourier component representing the Ir(110) row periodicity.

Additional DFT calculations were performed to obtain information on the interaction between the MnBr$_2$ and the Gr/Ir(110) substrate. Forced by computational limitations, we designed a smaller supercell that still correctly describes the chemistry at the MnBr$_2$ and Gr/Ir(110) interface (see Figure~S5 and Note 5 of the SI).
Overall, the MnBr$_2$ is weakly physisorbed to Gr/Ir(110) with an adsorption energy of $14$\,meV/\AA$^2$. No hybridization of MnBr$_2$ with its substrate is identified, but a small charge transfer from the substrate to the contact Br atoms is found. It is therefore plausible that a lateral modulation of the charge transfer is of relevance for the formation of the virtual MnBr$_2$/Ir(110) moiré.

The imprint of the Ir(110) row periodicity on the Gr/Ir(110) moiré and the formation of the virtual MnBr$_2$/Ir(110) moiré are unrelated to the well documented transparency of Gr on weakly interacting noble metal substrates. On Au(111) \cite{Tesch16}, Ag(111) \cite{Jolie15,Tesch16}, and Cu(111) \cite{Gonzalez16} tunneling conditions could be tuned to detect either quasiparticle scattering of metal surface state electrons or Gr and its moiré. The apparent transparency of Gr was explained by the slower decay of the surface state compared to the Gr local density of states into the vacuum \cite{Gonzalez16}. 

However, in the present case Gr on Ir(110) is strongly interacting with the substrate, rather than weakly. This is already evident from the fact that the Gr cover lifts the nanofacet reconstruction of Ir(110) \cite{Koch1991}. ARPES and DFT calculations prove the absence of a Dirac cone \cite{Kraus22} and thus a substantial inhomogeneity of the binding of C atoms to the substrate. The imprint of the spacing of the rows of the Ir(110) top layer on Gr is due to the fact that C atoms above the Ir(110) rows are strongly hybridized with the substrate, while those above the furrows interact much weaker with Gr. The chemical inhomogeneity of the C atoms in Gr/Ir(110) was already used to pattern adsorption \cite{Kraus22}. It is this inhomogeneity, that enters that enters the formation of the virtual MnBr$_2$/Ir(110) moiré and the unconventional structure of the MnBr$_2$/Gr/Ir(110) super-moiré.

\section{Conclusions}
 At 400\,K growth, MnBr$_2$ is nearly epitaxially aligned to Gr with densely packed rows rotated by 30$^\circ$, while at lower temperatures the orientation of MnBr$_2$ islands scatters more around the preferred epitaxial relation, as typical in van der Waals epitaxy. MnBr$_2$ exhibits a hexagonal lattice with a lattice parameter of 3.90 ± 0.01 Å, consistent with bulk 1T-MnBr$_2$. The MnBr$_2$ island morphology evolves from fractal to compact dendritic-skeletal islands as temperature increases. Nucleation is homogeneous at low temperatures but becomes heterogeneous at defects and step edges at room temperature and above. Second-layer nucleation is more probable than on bare Gr, while step-edge barriers are insignificant. Upon annealing, sublimation begins at 570\,K and MnBr$_2$ is fully desorbed at 670\,K. A comparative growth experiment suggests for single-layer MnBr$_2$ a similar growth behavior and structure on Gr/Ir(111). STS reveals a band gap for single-layer MnBr$_2$ of approximately 4.4\,eV with the valence band at $\approx$ -2.5\,eV and conduction band at $\approx$ +1.9\,eV. When tunneling into the MnBr$_2$ conduction band, the apparent STM height reaches 5.95\,\AA, almost the bulk interlayer spacing, while in the band gap it remains rather constant at the low value of 4\,\AA.  

A complex super-moiré arises from the interplay of MnBr$_2$, Gr, and Ir(110). Its strongest Fourier component corresponds to a virtual moiré between MnBr$_2$ and Ir(110), although the two lattices are not in contact. In real space this virtual moiré causes a dominant 2.7\,nm stripe pattern. Moreover, the linear combinations of these moiré wave vectors with the ones of the Gr/Ir(110) moiré with a 3.3\,nm stripe periodicity give rise to a super-moiré component with a 13.5\,nm beating pattern. The same virtual MnBr$_2$/Ir(110) moiré also enters the second super-moiré component due to linear combinations of its wave vectors with those of the MnBr$_2$/Gr moiré, which gives rise to orthogonal stripe segments in real space.

The analysis of the bias dependence of the Fourier components makes plain that the super-moiré is present only in the band gap of MnBr$_2$ when tunneling is from or to the MnBr$_2$/Gr interface, while when tunneling is into the conduction band the corrugation of MnBr$_2$ just reflects the corrugation of the Gr/Ir(110) moiré. Comparative experiments for MnBr$_2$ on Gr/Ir(111) display the same behavior of a geometric moiré when tunneling into the conduction band and an interface super-moiré present in the band gap. However, a virtual moiré is present only in the MnBr$_2$/Gr/Ir(110) case. Its existence is traced back to the inhomogeneous chemical binding of Gr to Ir(110), which causes an electronic and chemical modulation of Gr with 3.84\,\AA~periodicity of the dense-packed Ir(110) top layer rows, and thus also a modulation in the interaction with MnBr$_2$.

Overall, this comprehensive characterization of an example 2D TMDH will be useful for understanding the growth, structure, and super-moirés of other 2D materials in this class. The first characterization of a super-moiré composed of lattices with different symmetries broadened our understanding of how super-moirés can emerge. In addition, virtual moirés of lattices that are not in contact with each other may be of primary relevance in the formation of super-moirés. Since single-layer MnBr$_2$ is expected to host magnetic order and polarons, it will be interesting to explore whether any of these are affected by the super-moiré of MnBr$_2$/Gr/Ir(110).  

\section{Methods}

The experiments were carried out in two ultrahigh vacuum systems with base pressures below $1 \times 10^{-10}$\,mbar. Each system was equipped with standard MBE growth facilities, LEED or microchannel plate LEED (MCP-LEED), and STM operated at temperatures of 300\,K, 77\,K, and 1.7\,K. 

Ir(110) is cleaned and prepared in its unreconstructed state by cycles of 4.5\,keV Xe$^+$ ion sputtering, flash annealing to 1510\,K, and subsequent cooling to 400\,K in $1\times10^{-7}$\,mbar oxygen pressure. Ir(111) is cleaned by cycles of 1\,keV Ar$^+$ sputtering and subsequent flash annealing to 1510\,K. 

Single-crystal Gr was grown by heating cleaned unreconstructed Ir(110) to 1510\,K and exposing it to $3\times10^{-7}$\,mbar ethylene for 240\,s \cite{Kraus22}. On Ir(111), Gr was grown by exposing the clean Ir(111) to $1\times10^{-7}$\,mbar ethylene at 300\,K for 120\,s, flash annealing to 1470\,K without ethylene, and again exposure to $3\times10^{-7}$\,mbar ethylene for 600\,s at 1370\,K \cite{vanGastel09}. The quality of the as-grown Gr/Ir(110) or Gr/Ir(111) is checked by LEED and STM.

MnBr$_{2}$ is grown by sublimation of MnBr$_{2}$ molecules from MnBr$_{2}$ powder in a Knudsen cell heated to 670\,K. The Knudsen cell was placed 8\,cm from the sample, producing a deposition rate of $3 \times 10^{-4}$\,ML/s. Here, a monolayer (ML) corresponds to complete surface coverage by a single layer of MnBr$_{2}$, equivalent to $7.6 \times 10^{18}$\,MnBr$_{2}$ molecules m$^{-2}$ s$^{-1}$. 

LEED patterns were acquired at 300\,K using electron energies ranging from 90 to 150\,eV. For MCP-LEED measurements, distortions in the reciprocal space due to the flat microchannel plate geometry were corrected using a Python script.    

STM was performed at 300\,K in a variable temperature and at 77\,K and 1.7\,K in a bath cryostat STM system. Constant-current topographies were recorded with sample bias $V_\mathrm{b}$ and tunneling current $I_\mathrm{t}$, as detailed in the respective figure captions. The STM images were analyzed and processed (plane subtraction, contrast correction) using WSxM software \cite{Horcas2007}. 
STS spectra were obtained at 1.7\,K with stabilization bias $V_{st}$ and stabilization current $I_{st}$ using the standard lock-in technique with modulation frequency $f_{mod}$ and modulation voltage $V_{mod}$, as specified in the captions.

Our \textit{ab initio} density functional theory 
(DFT) \cite{PR136_B864,PR140_A1133} calculations were carried 
out using the projector augmented wave (PAW) method \cite{PRB50_17953} 
as implemented in the Vienna Ab initio Simulation Package (VASP) \cite{PRB47_558,PRB54_11169,PRB59_1758}. To account for the van der Waals interactions in the Gr/Ir(110) and MnBr$_2$/Gr/Ir(110) systems, we employed the non-local vdW-DF2 correlation energy functional \cite{PRB82_081101R} combined with a re-optimized B86b exchange functional \cite{PRB89_121103R, JCP86_7184}. 
Gr on Ir(110) was modeled using a slab comprising three Ir layers and an in-plane unit cell of $\approx 32.86 \times 30.12 $\,\AA~amounting to 614 atoms (350 C and 264 Ir). The MnBr$_2$/Gr/Ir(110) system was simulated by using an in-plane unit cell of $\approx 7.04 \times 28.46 $\,\AA~that contains 192 atoms (i.e., 14 Mn, 28 Br, 84 C and 66 Ir).
The ground-state geometry and electronic structure were determined using a plane-wave kinetic energy cutoff of 500\,eV and a threshold for Hellmann-Feynman forces of $\approx 0.005$\,eV/\AA.

\section{acknowledgments}
	\noindent
   Funding from the Deutsche Forschungsgemeinschaft (DFG) through CRC 1238 (project number 277146847, subprojects A01, C01, and B06) is acknowledged. J.F. acknowledges financial support from the DFG through project FI 2624/1-1 (project no. 462692705) within SPP 2137. W.J. acknowledges financial support by the DFG priority program SPP2244 (project no. 535290457). The authors acknowledge the computing time granted by the JARA Vergabegremium and provided on the JARA Partition part of the supercomputer JURECA at Forschungszentrum Jülich. 

\section*{Conflict of Interest}
\noindent
The authors declare no conflicts of interest.

\section*{Data Availability Statement}
\noindent
The data that support the findings of this study are available from the
corresponding author, upon reasonable request.

\bibliographystyle{apsrev4-2}
\bibliography{Ref}

\newpage

%\date{\toda
\maketitle
\section{Supplementary Information}
\setcounter{figure}{0} % Resets counter to 0 for this section
%\subsection*{Supplementary Note 1. Apparent height of
%  \texorpdfstring{$\mathrm{MnBr_2}$}{MnBr2}}
%\subsection*{Supplementary Note 1. Apparent height of  MnBr$_2$}

%\subsection*{Supplementary Note 1. Apparent height of MnBr\textsubscript{2}}

\subsection*{%
  \texorpdfstring
    {Supplementary Note 1. Apparent height of MnBr\textsubscript{2}}%
    {Supplementary Note 1. Apparent height of MnBr2}%
}
\begin{figure}[hbt!]
\includegraphics[width=\textwidth]{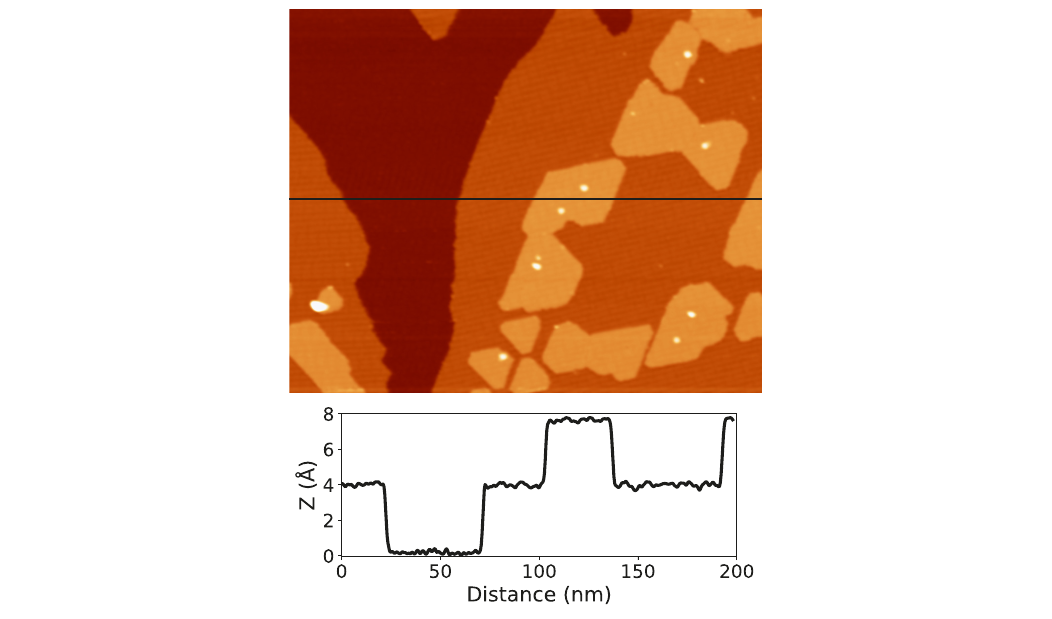}
\caption{STM topograph of MnBr$_2$ on Gr/Ir(110). Lower panel shows the height profile along the black line in the STM topograph. STM image is obtained at 300\,K with (b) $V_\mathrm{b}$ = -2\,V, 50\,pA. Image size: 200\,nm $\times$ 160\,nm } 
  \label{Sfig_heights} 
\end{figure}

\newpage
\subsection*{Supplementary Note 2. Comparison of single layer MnBr$_2$ grown on Gr/Ir(111) and Gr/Ir(110)}
\begin{figure}[hbt!]
\includegraphics[width=\textwidth]{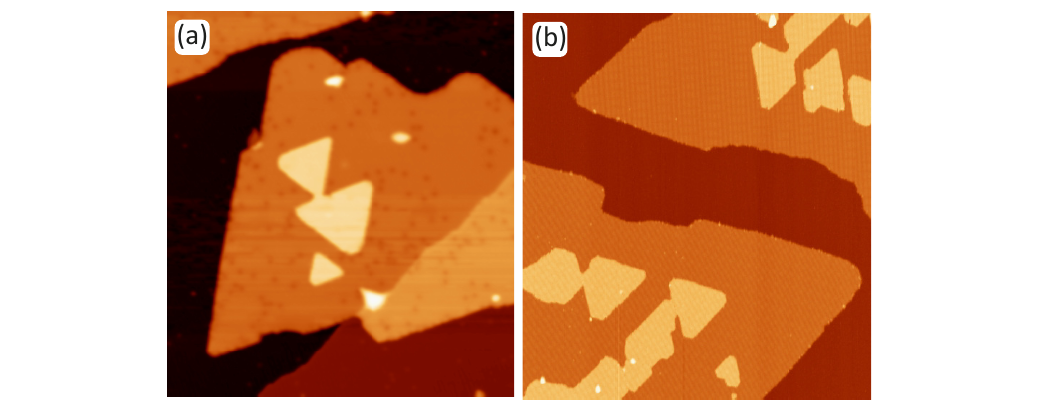}
\caption{STM topograph of MnBr$_2$ grown at 400\,K on (a) Gr/Ir(111) and (b) Gr(Ir(110). (b) is identical to Figure 1(b) of the main manuscript and is reproduced here for direct comparison. Tunneling parameters are for (a) $V_\mathrm{b}$ = 3.5\,V, 20\,pA. and for (b) $V_\mathrm{b}$ = -2\,V, 50\,pA. Image size for (a) and (b) is 180\,nm $\times$ 200\,nm.}
  \label{Sfig_heights} 
\end{figure}

\newpage
\maketitle
\subsection*{Supplementary Note 3. Comparison of Gr/Ir(110) with MnBr$_2$/Gr/Ir(110)}
\begin{figure}[hbt!]
\includegraphics[width=\textwidth]{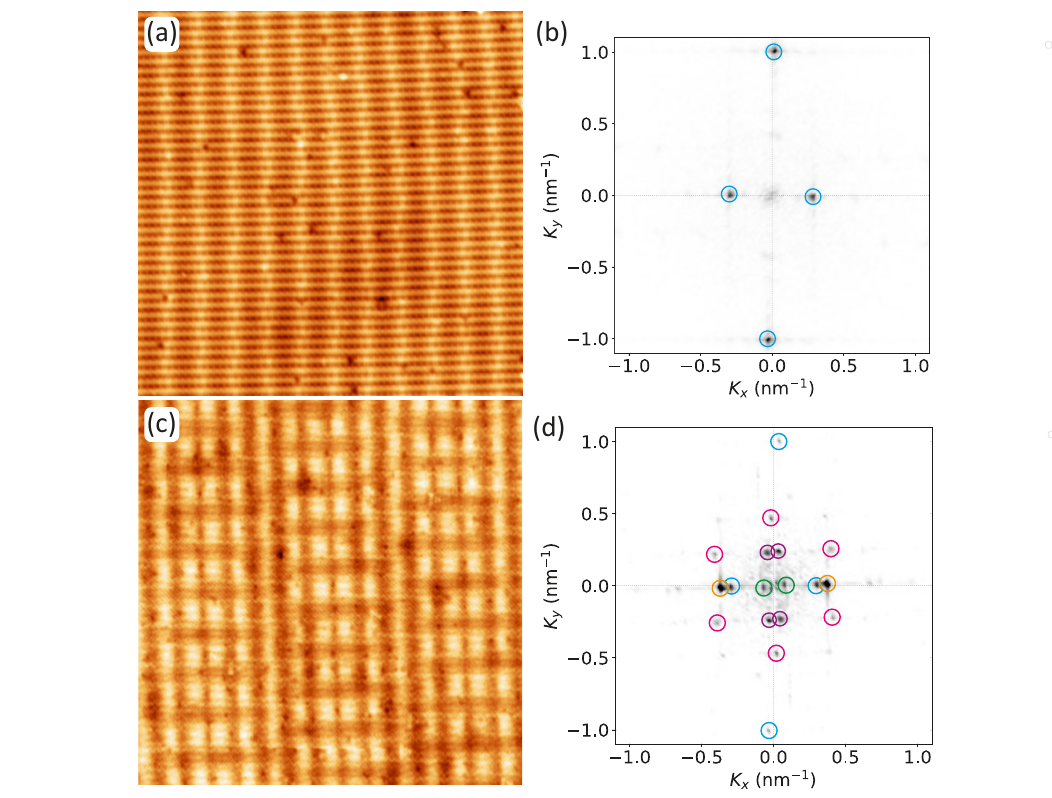}
\caption{(a) Moiré resolution STM topography of Gr/Ir(110) and (b) its corresponding FFT. (c) Moiré resolution STM topography of MnBr$_2$/Gr/Ir(110) and (c) its corresponding FFT. The Fourier components encircled blue in (b) and (d) are identical. (c) and (d) are Figures~5(a) and (b) of the paper and are reproduced for direct comparison. STM imaging parameters are (a) $V_\mathrm{b}$ = -1\,V, 100\,pA and (c) $V_\mathrm{b}$ = -1\,V, 50\,pA. STM image size is 40\,nm $\times$ 40\,nm.} 
  \label{Sfig_heights} 
\end{figure}

\newpage
\maketitle
\subsection*{Supplementary Note 4. Contour of constant total charge density above Gr/Ir(110)}
\begin{figure}[hbt!]
\includegraphics[width=\textwidth]{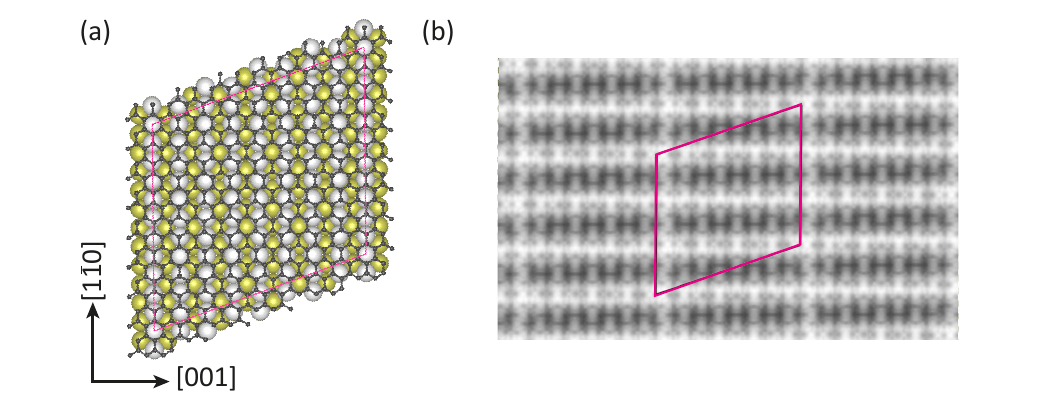}
\caption{(a)  Ball model of atomic arrangement in the supercell. Green gold balls: bottom-layer Ir atoms; white balls: top-layer Ir atoms; dark gray balls: C atoms. A pink rhombus indicates the supercell size in the DFT calculations. (b) The total charge density plotted above the surface for an isosurface value of $2.1 \times 10^{-10}$\,electrons/\AA$^3$.
The grayscale encodes the variation in isosurface height. The periodicity of the Ir(110) top layer substrate rows with their 3.84\,\AA~spacing is prominent.
}
\label{Sfig_charge} 
\end{figure}

\newpage
\maketitle
\subsection*{Supplementary Note 5. DFT simulation of MnBr$_2$/Gr/Ir(110)}
\begin{figure}[hbt!]
\includegraphics[width=\textwidth]{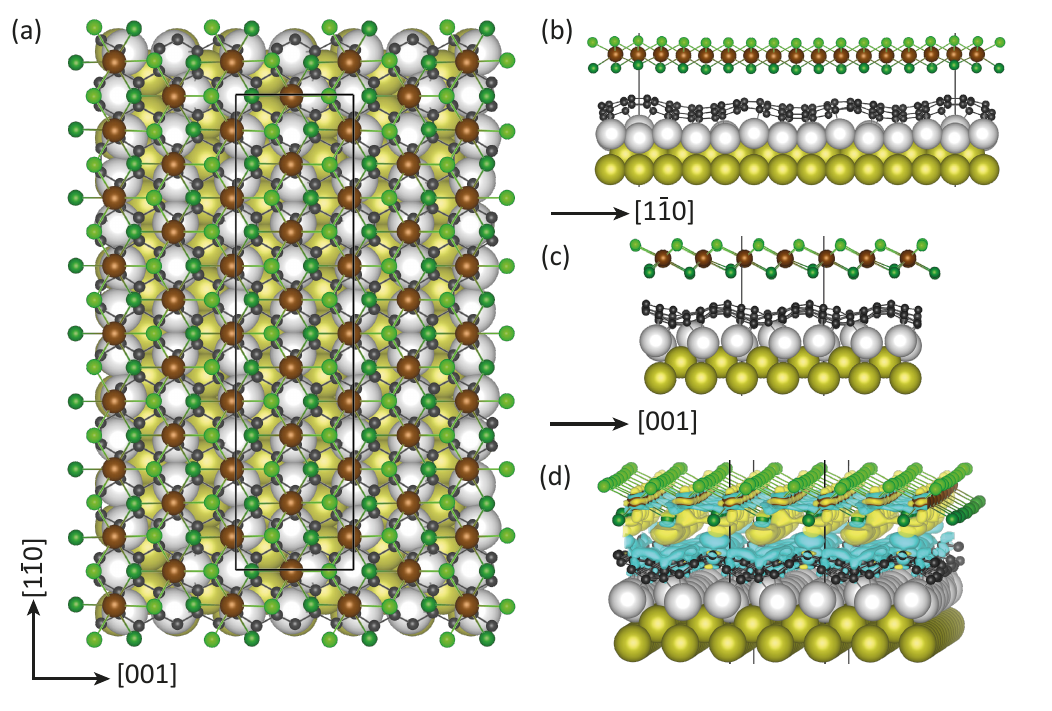}
\caption{(a) Top view of the relaxed structure of MnBr$_2$ on Gr/Ir(110). The size of supercell is indicated by black rectangle. It is smaller by a factor of $\approx 4$ compared to the supercell used to model the Gr/Ir(110) moiré in Figure~6(d) of the main paper. (b),(c) Side views along (b) $[001]$ direction and (c) $[1\Bar{1}0]$ direction. Black lines indicate boundaries of supercell. (d) The charge density difference is plotted above the surface for an isosurface value $4.9 \times 10^{-8}$\,e/\AA$^3$. Yellow indicates charge density accumulation and blue charge density depletion.}
\label{Sfig_DFT} 
\end{figure}

DFT calculations for the entire super-moiré unit cell are impossible. Even including the MnBr$_2$ layer in the unit cell as shown in Figure 9(a) is computationally very expensive and practically not feasible. However, to get an insight into the interaction that occurs between MnBr$_2$ and the Gr/Ir(110) substrate we can use a smaller supercell that correctly captures the chemistry at the interface of MnBr$_2$ and Gr/Ir(110).
The supercell used in our calculations is indicated by a black rectangle in Figure~S5(a) and contains 14 Mn, 28 Br, 84 C, and 66 Ir atoms. The relative lattice parameters with respect to their nominal values are 0.92 for Ir, 0.96 for Gr, and 1.05 for Mn. We note here that the choice of the unit cell implies a compressive stress for Gr/Ir(110) substrate and tensile stress for the MnBr$_2$ layer.
\\
Our calculations show that the distances between the C atoms of Gr and the lower Br atoms of MnBr$_2$ vary between 2.9\,\AA~and 4.3\,\AA, as is typical for van der Waals binding [see Figures~S5(b) and (c)]. The adsorption energy for the unit cell used in the calculation is 2.86\,eV, corresponding to 14 meV$/\AA^2$ or 34\,meV per carbon atom. 
As a note, the interaction occurring between the graphene layers in graphite is 61\,meV per carbon atom \cite{Zacharia2004}. Furthermore, we found that there is practically no hybridization between Br and C atoms. We also investigated the charge transfer that occurs at the interface. Figures~S5(d) plots the charge density difference when the two parts of the system, Gr/Ir(110) and MnBr$_2$, are joined. Isosurfaces for charge accumulation and charge depletion are colored yellow and blue, respectively. For the entire unit cell 0.092 of an electron is transferred from graphene to MnBr$_2$, mainly to the lower bromine layer in contact with Gr. Therefore, we can also rule out any significant ionic contribution to the binding between MnBr$_2$ and Gr/Ir(110).
As a consequence, our calculations demonstrate that the MnBr$_2$ binds the Gr/Ir(110) surface only via rather weak van der Waals interactions.

\end{document}